\documentclass[twocolumn,showpacs,preprintnumbers,amsmath,amssymb]{revtex4}

\usepackage{graphicx}
\usepackage{dcolumn}
\usepackage{bm}

\begin{document}

\title{Unveiling the tachyon dynamics in the Carrollian limit}

\author{C. Escamilla-Rivera}
 \email{celia_escamilla@ehu.es}
 \affiliation{Fisika Teorikoaren eta Zientziaren Historia Saila, Zientzia 
eta Teknologia Fakultatea, Euskal Herriko Unibertsitatea, 644 Posta Kutxatila, 48080, Bilbao, Spain.}

\author{G. Garcia-Jimenez}
 \email{ggarcia@fcfm.buap.mx}
 \affiliation{Facultad de Ciencia Fisico Matematicas de la Universidad Autonoma de Puebla, P.O. Box 1364, 72000, Puebla, Mexico.}
        
\author{O. Obregon}
 \email{obregon@fisica.ugto.mx}
 \affiliation{Departamento de Fisica de la Universidad de
    Guanajuato, C.P 37150, Leon, Guanajuato, Mexico.}
    
\date{\today}

\begin{abstract}
We briefly study the dynamics at classical level of the Carrollian limit, with vanishing speed of light and no possible propagation of signals, for a simply effective action in a flat space with a open string tachyon as scalar field. The canonical analysis of the theory indicates that the equation of motion is of Dirac type contrary to non-relativistic case where the equation is of Schrodinger type. The ultimate intention is to analize the latter case with electromagnetic fluxes finding that in this case the open string tachyon cannot be interpreted as time.
\end{abstract}

         
\maketitle

\section{Introduction}
In past years the role of the tachyon in certain string theories has been explored and this has resulted in a better understanding of the D-brane decaying process \cite{Sen:2002qa},\cite{Gorini:2003wa}. The basic idea is that the usual open string vacuum is unstable, but there exists a stable vacuum with zero energy density which is stable, which a tachyon field $T(x)$ naturally moves to. Nevertheless it seems that aspects of this process can be compared with a simple effective field theory models. In this case maybe the simplest model was proposed by Sen \cite{Sen:2002qa}. This success of effective action methods, together with the difficulties of other approaches described encourages one to pursue this further and to attempt a exact description of the cosmology of tachyon rolling \cite{Gibbons:2002md}.

Moreover, in the case where there are electromagnetic fluxes, the tachyon field is on the same footing as a transverse scalar in the Dirac-Born-Infeld action for a brane \cite{Gibbons:2002tv}. In this case we look for a solution with a constant electromagnetic field and find that the condensed state at $V(T)\rightarrow 0$ is given by ${\dot{T}}^{2}+{E}^{2}=1$, where $\dot{T}$ means derivative with respect to the dimensionless time of tachyon field and $E=|\vec{E}|$. To understand the dynamics it is convenient to do the Hamiltonian formulation of the theory.

The present manuscript is organized as follows. In Sec. II we review the role of the open string tachyon in the field theory and how this scalar field takes place in the decaying process.

In Sec. III we describe what we have called the Carrollian limit mechanism for open string states. Since this entails familiarity with Carroll group,
I planned to include also the Galilean group and the differences between them.

In Sec. IV we discuss some aspects when this theory is coupled to gravity.

In Sec. V we use the low energy effective action of the open string tachyon and take the two possible limits: first the Galileo limit (when $c\rightarrow \infty$), i.e the contravariant metric $\eta^{\mu \nu}=(-c^{-2},1,1,1$) is well defined, contrary to the Carrollian limit (when $c\rightarrow 0$). For this case we obtain in the Hamiltonian formulation a Dirac type equation.

In section VI we use again the effective action and consider the case in which $F_{\mu\nu}\neq 0$ to find that the tachyon is accelerated and emits radiation in the direction of the electromagnetic field.

\section{Open string tachyon in field theory}
To understand clearly the tachyon dynamics we take into consideration a real scalar field $\phi$ in a flat space-time. The Lagrangian of this theory is given by
\begin{equation}
\begin{aligned}
	L=-\frac{1}{2}{\left(\partial\phi\right)}^{2}+\frac{1}{2}V\left(\phi\right),
\end{aligned}
\end{equation}
where $V\left(\phi\right)$ is the scalar field potential. In perturbation theory usually we expand the potential of the form
\begin{equation}
\begin{aligned}
V\left(\phi\right)=V_{0}+{\lambda}_{1}\phi+{\lambda}_{2}{\phi}^{2}+{\lambda}_{3}{\phi}^{3}+\ldots,
\end{aligned}
\end{equation}
and assume that ${\phi}^{n}$ becomes small, as the system evolves in time for large $n$. We also know that ${\lambda}_{2}=m^{2}$, i.e this is a mass term. In this expansion we have two interesting cases: a) $V''\left(\phi=0\right)={\lambda}_{2}>0$, i.e ${m}^{2}>0$, the theory has a real mass spectrum. In this case, the solutions of ${\phi}^{n}$ decrease for large $n$ over time and therefore the perturbation theory is valid. And, b) $V''\left(\phi=0\right)={\lambda}_{2}<0$, i.e ${m}^{2}<0$, and the theory has a imaginary mass spectrum, i.e a tachyon. In this case, the solutions of ${\phi}^{n}$ grow to infinity for large $n$ over time and as a consequence the perturbation theory is no longer valid. The latter case indicates that the theory is unstable around $\phi=0$. The usual way to solve this is to find a critical point (stable) $\phi={\phi}_{0}$, where the perturbation theory must be valid and obtain a real mass spectrum. Sen found a clear way to study the tachyons in certain string theories similar to previous case \cite{Sen:2002qa}. He suggested that at a effective theory level (low energy) the tachyons indicate the instability of the system and correspond to decaying processes systems of open string with branes. 

If we configure the system to a initial time so that the tachyon have a initial amplitude in $T=0$ we have an unstable state and $V\left(T=0\right)>0$. Any small perturbation would allow to the tachyon potential descends and reach any of the two minimum. In theory these two minimum are stable under small perturbations in the field $\phi$ and its mass spectrum is real.

On the other hand, as proposed by Sen, Gibbons suggested analyze the coupling to gravity and considering the resulting cosmology \cite{Gibbons:2002md},\cite{Gibbons:2003gb}. In this case he found that Sen's action is defined with a covariant metric $\eta_{\mu\nu}$ and then the limit of the theory is correct when $c\rightarrow 0$, because in this case there exists a regular metric, this limit is the so-called Carrollian limit. Of course, Gibbons took this into consideration for classical cases and geometric level (collapse of cones of light).

\section{How does the Carrollian limit work?}
The Carroll limit is defined as the limit when $c\rightarrow 0$, where $c$, as we know, is the speed of light, which in this context is seen as a parameter. In this limit, the resulting space is called Carroll space-time and the symmetries of this space define a transformation group called the Carroll group. Then, given a theory that incorporates the speed of light as a parameter (i.e a relativistic theory) it is possible to make this limit, also called \textit{contraction}, and obtain new properties very different from what we originally had. A well known example of this contraction is the case of the Poincare group, in which it is possible to get the Galilean group through the limit $c\rightarrow\infty$. The latter limit is physically interpreted as the unreal limit of the theory. However, from a geometric point of view, we can see that given the line element
\begin{equation}\label{line element}
\begin{aligned}
	ds^{2}=-c^{2}dt^{2}+dx_{i}^{2},
\end{aligned}
\end{equation}
we introduce the covariant metric
\begin{equation}\label{covariant}
\begin{aligned}
	\eta_{\mu\nu}=(-c^{2},1,1,1),
\end{aligned}
\end{equation}
where $ds^{2}=\eta_{\mu\nu}dx^{\mu}dx^{\nu}$. The inverse matrix is just the contravariant matrix
\begin{equation}\label{contravariant}
\begin{aligned}
	\eta^{\mu\nu}=(-c^{-2},1,1,1).
\end{aligned}
\end{equation}
The remarkable thing is that in the limit $c\rightarrow \infty$, the contravariant metric (\ref{contravariant}) is well defined and the covariant not, while in the limit $c\rightarrow 0$, the opposite happens. The first case defines a structure called the Newton-Cartan and the second defines a Carroll space-time.

\section{Coupling to gravity}
Follow the common wisdom and assume that the relavant action in a flat space is
\begin{equation}
\begin{aligned}
	S=\int{d^{4}x L},
\end{aligned}
\end{equation}
where the Lagrangian density has the form of Born-Infeld
\begin{equation}
\begin{aligned}
	L=-V(T)\sqrt{-det A_{\mu\nu}},  \quad  A_{\mu\nu}=\eta_{\mu\nu}+\partial_{\mu}\partial_{\nu}T.
\end{aligned}
\end{equation}
With this in mind, the natural way to introduce the gravitational field is by hand,
\begin{equation}
\begin{aligned}
	S=-\int{d^{4}x V(T) \sqrt{-g} \sqrt{1+g^{\mu\nu}\partial_{\mu}T\partial_{\nu}T}}.
\end{aligned}
\end{equation}
The term inside the root is the metric associated to the open string sector (i.e only tachyonic matter)
\begin{equation}
\begin{aligned}
 G_{\mu\nu}=g_{\mu\nu} +\partial_{\mu}T \partial_{\nu}T.
\end{aligned}
\end{equation}
For the case when the open string tachyon $T$ depends on time,
\begin{equation}\label{gmetric}
\begin{aligned}
	G_{\mu\nu}= diag(-1+{\dot{T}}^{2},1,1,1).
\end{aligned}
\end{equation}
As explained before, the tachyon condensate takes place in the limit when its velocity tends to one, so (\ref{gmetric}) can be rewritten as
\begin{equation}
\begin{aligned}
	G_{\mu\nu}\rightarrow diag (0,1,1,1).
\end{aligned}
\end{equation}
Here the covariant metric is well defined and therefore the tachyon condensate naturally gives us a Carroll spacetime.

\section{Sen's action in two limits}
The open string tachyon can be described by an effective action where the flat space has a Lagrangian given by
\begin{equation}\label{tachyon lagrangian}
\begin{aligned}
	L=-V(T)\sqrt{-det(\eta_{\mu\nu}+\partial_{\mu}T\partial_{\nu}T)},
\end{aligned}
\end{equation}
where $V(T)$ is the tachyon potential and has a positive maximum at the origin and a minimum at $T=T_{0}$. At this point the potential vanishes. The equation (\ref{tachyon lagrangian}) reproduces correctly the asymptotic behaviour $T\rightarrow \pm \infty$ for the energy density and pressure obtain by Sen, and therefore, it is a good model to describe the effective theory \cite{Gibbons:2002md}.

For a homogeneous tachyon $T=T(t)$ the equation (\ref{tachyon lagrangian}) has the form
\begin{equation}\label{tachyon lagrangian2}
\begin{aligned}
	L=-cV(T)\sqrt{1-\frac{{\dot{T}}^{2}}{c^{2}}}.
\end{aligned}
\end{equation}
We known that for the Galileo group $c\rightarrow \infty$, if we expand the root and take this limit our Lagrangian can be written as
\begin{equation}\label{tachyon lagrangian3.1}
\begin{aligned}
	L\cong -cV(T)\left(1-\frac{{\dot{T}}^{2}}{2c^{2}}+\ldots \right).
\end{aligned}
\end{equation}

On the other hand, if we take the Carrollian limit $c\rightarrow 0$ over (\ref{tachyon lagrangian2}) the expansion of the Lagrangian is now 
\begin{equation}\label{tachyon lagrangian3}
\begin{aligned}
	L\cong -iV(T)\dot{T}+\ldots.
\end{aligned}
\end{equation}
Calculating the canonical momentum associated with the tachyon we get
\begin{equation}
\begin{aligned}
	\Pi_{T}\equiv \frac{\partial L}{\partial \dot{T}}=-iV(T).
\end{aligned}
\end{equation}
From here we have the following constraint
\begin{equation}\label{constraint}
\begin{aligned}
	\Phi_{T}=\Pi_{T}+iV(T)\approx 0,
\end{aligned}
\end{equation}
where the notation $\approx$ means weakly zero in Dirac's language. Due to the constraint (\ref{constraint}), the canonical Hamiltonian is zero and therefore the Hamiltonian of the theory (total Hamiltonian) is given by the product of an arbitrary function (Lagrange's multiplier) and the constraint (\ref{constraint}). If we impose at quantum level the total Hamiltonian and therefore the constraint then we see that the Hamiltonian is not hermitian if the potential is real. So, the dynamics are defined at quantum level only if the potential is pure imaginary, i.e when exists the creation of tachyons. At classical level, the solution for the tachyon is formally given by the temporal integral of the Lagrange's multiplier. As we can easily see by calculating the equation of Hamilton using the total Hamiltonian we can interpret the tachyon as time only if the Lagrange's multiplier is a constant.

\section{Inclusion of fluxes}
We now turn our attention to the case in which $F_{\mu\nu}\neq 0$. Gibbons got the following Lagrangian for the tachyon condensation $V(T)\rightarrow 0$,
\begin{equation}
\begin{aligned}
L=-V(T)\sqrt{-det(\eta_{\mu\nu}+{\partial}_{\mu}T{\partial}_{\nu}T)+F_{\mu\nu}}.
\end{aligned}
\end{equation}
In this last equation we only added to equation (\ref{tachyon lagrangian}) the electromagnetic term. If $E=|\vec{E}|$ is a constant, then ${\dot{T}}^{2}+{\vec{E}}^{2}\rightarrow 1$, when $T\rightarrow \infty$. In the literature we only found the effects of the electric field $\vec{E}$. In the same line, our intention is to discuss what happens in the case when the magnetic field $\vec{B}$ exists? The above expresion changes in the following way: the matrix obtain for this case is
\begin{equation}
\begin{aligned}
G_{\mu\nu}=\left(\begin{array}{cccc}
-1+\frac{{\dot{T}}^{2}}{c^{2}}  &  \lambda E_{1}   &   \lambda E_{2}   &  \lambda E_{3}	\\
-\lambda E_{1}                  &  1               &   \lambda cB_{3}  &  -\lambda cB_{2} \\   
-\lambda E_{2}                  &  -\lambda cB_{3} &   1               &  \lambda cB_{1} \\
-\lambda B_{3}                  &  \lambda cB_{2}  &   -\lambda cB_{1} &  1 
\end{array} \right),
\end{aligned}
\end{equation}
where ${\lambda}^{2} =c^{-2}$. After lengthy but otherwise straightforward calculations, we can write the Lagrangian
\begin{equation}\label{tachyon lagrangian4}
\begin{aligned}
L=-V(T)\sqrt{1-\frac{{\dot{T}}^{2}\left(1+c^{2}{\vec{B}}^{2}\right)}{c^{2}}+{\vec{B}}^{2}-\frac{{\vec{E}}^{2}+{\left(\vec{E}\cdot \vec{B}\right)}^{2}}{c^{2}}}.
\end{aligned}
\end{equation}
In the limit of tachyon condensation we have
\begin{equation}\label{tachyon lagrangian5}
\begin{aligned}
{\dot{T}}^{2}\left(1+{\vec{B}}^{2}\right)+\left({\vec{E}}^{2}-{\vec{B}}^{2}\right)+{\left(\vec{E}\cdot \vec{B}\right)}^{2}=1,
\end{aligned}
\end{equation}
where we consider $c=1$ for simplicity. In the case proposed by Gibbons ($\vec{B}\approx 0$) we note that the tachyon is accelerated and therefore emits radiation and the propagation is in the direction of the electric field $\vec{E}$. The allowed range for tachyon velocity is then $\left[0,\sqrt{1-\vec{E}^{2}}\right]$.

In our case, the propagation of radiation occurs in the component of electromagnetic field, but if we consider $\vec{E}\approx 0$, may imply that $\dot{T}\rightarrow 1$, in other words, the condensate is not affected in the presence of magnetic fields. This suggests that the tachyon does not interact with this field. 

It should be remarkable, however, that under the presence of a uniform electromagnetic field, the open string tachyon cannot be interpreted as time in the sense of a Schrodinger type equation \cite{GarciaCompean:2005zn} because as we can see from (\ref{tachyon lagrangian5}) the tachyon does not decouple from the electromagnetic field.

\begin{acknowledgments}
C. Escamilla-Rivera would like to thank J. Socorro-Garcia for very helpful discussions. This work was supported by CONACyT, Fundacion Pablo Garcia and FUNDEC Mexico. 
\end{acknowledgments}


\begin{thebibliography}{9}                                                                                 
\bibitem{Sen:2002qa} A. Sen, \textit{Time and Tachyon}, Int. J. Mod. Phys. A18(2003) 4869-4888, hep-th/0209122.

\bibitem{Gorini:2003wa} V. Gorinni, A. Kamenshchik, U. Moschella, V. Pasquier \textit{Tachyons, Scalar Fields and Cosmology}, Phys. Rev. D69 (2004)123512, hep-th/0311111.

\bibitem{Gibbons:2002md} G.W. Gibbons \textit{Cosmological Evolution of the Rolling Tachyon}, arXiv:hep-th/0204008v2 (2002).

\bibitem{Gibbons:2003gb} G.W. Gibbons \textit{Thoughts on Tachyon Cosmology}, arXiv:hep-th/0301117v1 (2003).

\bibitem{GarciaCompean:2005zn} H. García Compeán, G. García Jiménez, O. Obregón and C. Ramírez \textit{Tachyon driven quantum cosmology in string theory}, Phys. Rev. D 71, 063517 (2005).

\bibitem{Gibbons:2002tv} Gibbons, Gary and Hashimoto, Koji and Yi, Piljin \textit{Tachyon condensates, Carrollian contraction of Lorentz
group, and fundamental strings}, hep-th/0209034 (2002).

\end{thebibliography}
\end{document}